# Performance Analysis of Vertical Axis Wind Turbine Clusters: Effect of Inter-Turbine Spacing and Turbine Rotation

Dinesh Kumar Reddy G[1], Mayank Verma[1,2], Ashoke De[1,3 a)]

[1]*Department of Aerospace Engineering, Indian Institute of Technology Kanpur, 208016, Kanpur, India.*

[2]*IIHR – Hydroscience & Engineering, University of Iowa, Iowa City, IA 52242, USA.*

[3]*Department of Sustainable Energy Engineering, Indian Institute of Technology Kanpur, 208016, Kanpur, India.*

Wind energy has emerged as a viable alternative to fossil fuels, with vertical axis wind turbines (VAWTs) gaining popularity due to their efficiency and adaptability. Combining the Actuator Line Method (ALM) with Large Eddy Simulation (LES) enables accurate performance evaluations, facilitating the design and optimization of wind turbines. The present study invokes ALM-based methodology to perform calculations for the VAWTs. The results of the LES simulations of the VAWTs have been extensively validated against the available experimental and numerical data. The study further explores a VAWT cluster of three turbines by investigating the influence of turbine spacing (in both in-line and staggered configuration) on cluster performance. The study shows that the configuration with a streamwise separation ($X_{sep}$) of 0.34D and a transverse separation ($Y_{sep}$) of 2.5D exhibits superior performance to other combinations owing to increased kinetic energy in the wake for the downstream turbines. Further, we have presented the effect of varying the rotation direction (in combinations of Clockwise and Counter-Clockwise rotation) for the individual turbines in the 3-turbine cluster for the two configurations: in-line ($X_{sep}$ = 0D, $Y_{sep}$ = 2.5D) and staggered ($X_{sep}$ = 0.34D, $Y_{sep}$ = 2.5D). Staggered counter-rotating turbine cases show reduced performance compared to co-rotating cases, specifically, the clockwise co-rotating (C-C-C) configuration. In the in-line configuration, counter-rotating setups outperform co-rotating ones. Counter-rotation analysis reveals that reducing streamwise separation allows turbines to align in line without sacrificing performance, thereby increasing the power density of the turbine cluster.

## I. INTRODUCTION

The increasing global demand for cost-effective and clean energy has paved the way for research on various renewable energies, of which wind energy is the most popular due to its abundance. Horizontal axis wind turbines (HAWTs) have been at the forefront of wind energy research and applications because their power coefficient is much higher than that of VAWTs. The VAWTs are a promising alternative to the HAWTs, particularly in regions with limited wind resources or spatial constraints, and their design characteristics enable them to capture wind from various directions, making them adaptable to urban environments where wind direction is unpredictable. The wake dynamics of VAWTs are different from their HAWT counterparts because of their axis of rotation. The wake exhibits pronounced horizontal asymmetry attributed to a deep dynamic stall on the windward side and the Magnus effect. At the same time, it undergoes a slight downward shift in the vertical

___

a) Author to whom correspondence should be addressed.  Electronic mail:  ashoke@iitk.ac.in

direction. The primary mechanism responsible for re-energizing the wake is the presence of counter-rotating vortical structures, facilitating momentum recovery[1].

In modern wind farms, HAWTs are commonly positioned 3-5 rotor diameters apart in the cross-wind direction and 6-10 diameters apart in the streamwise direction to maximize power output while minimizing aerodynamic interference between adjacent turbines[2]. For realistic cost ratios (ratio of land surface costs and turbine costs), the optimal average turbine spacing in large HAWT wind farms is significantly higher (around 15D) compared to current wind farm implementations (around 7D)[3]. The traditional approach to wind turbine spacing in wind farms is primarily dictated by restricting wake-induced fatigue loads on turbines positioned downstream of preceding rows[4]. A VAWT pair has a wake recovery distance of 6D, which is way less than the 14D for the flow behind a single HAWT[5, 6]. While individual VAWTs exhibit lower performance than HAWTs, the collective performance of VAWTs in a grouped configuration is improved due to the flow acceleration between the VAWTs[7]. Consequently, by arranging Vertical Axis Wind Turbines (VAWTs) with narrower spacings, it is possible to achieve higher power densities (i.e., power per unit area utilized) than conventional HAWT wind farms[8].

The performance prediction and wake interactions between turbines in wind farms can be effectively estimated through full computational fluid dynamics (CFD) analysis, with accuracy depending on the chosen turbulence model[9-12]. Despite achieving full blade resolution and accurate evaluation of wake dynamics, the computational cost for analyzing large wind farms remains prohibitively high. Therefore, there is a demand for simplified models, known as reduced order models, that offer faster computation while maintaining reasonable accuracy. The selection of appropriate reduced order models, such as the vortex model, actuator disc model (ADM), and actuator line model (ALM), depends on the desired level of accuracy necessary for predicting turbine performance[13,14]. Despite its computing efficiency and widespread application in research and the industry, the actuator disc model is inaccurate in depicting a VAWT[15]. On the other hand, Sorensen et al.[15] developed ALM, which effectively monitors individual blade elements and, when coupled with large eddy simulation (LES), demonstrates the accurate analysis of wind farms.

The limited availability of open-source codes for implementing ALM in VAWT analysis is challenging. One such open-source code, turbinesFoam[16], has been developed by Bachant et al.[16] and is based on the OpenFOAM framework. Similarly, Bartholomew et al.[17]. developed an open-source Xcompact3D framework for modeling turbulent flow analysis, and the code has been extensively used for other studies[19-20]. In this study, we have invoked and modified the Xcompact3D[17] code as per our need to investigate VAWTs. Compared to HAWTs, the primary challenge in VAWTs is the number of actuator lines required for modeling. Also, VAWT modeling requires the incorporation of strut supports. Furthermore, implementing unsteady effects such as dynamic stall and added mass differs between HAWTs and VAWTs.

Hezaveh et al.[21] presented a novel optimization approach for large vertical axis wind turbine (VAWT) farms, which combines the Actuator Line Model (ALM) with Large Eddy Simulation (LES). They proposed an optimal architecture consisting of a triangular cluster design with three VAWTs, aiming to maximize flow acceleration and serve as a basis for large wind farms. The authors comprehensively evaluated wind farm performance, analyzing the inter-turbine distance within a cluster while considering factors such as total power generation, wind direction variability, and wake recovery. They discovered that the staggered configuration outperformed the conventional one by comparing staggered and regular layouts using the optimal spacing (L/D). This superiority is attributed to the utilization of flow acceleration facilitated by neighboring clusters in the staggered configuration, which enhances intra-cluster and inter-cluster synergy. In their analysis, Zhang et al.[22] investigated the performance of three turbine configurations (V shape, reverse V shape, and line shape) by varying the spacings in cross-stream and streamwise directions. They observed that the line configuration consistently outperformed the other shapes across all combinations of spacings.

Christie et al.[23] developed a simplified model to assess the maximum achievable power densities in different wind turbine array configurations. They investigated factors such as array size, turbine spacing, and setback distance (the distance between the outermost turbine and the boundary of the wind farm) and evaluated their impact on performance. The analysis used wake models focused on the wake effect when a turbine is positioned behind multiple turbines. Larger arrays demonstrated better outcomes as the setback distance increased. While the wake effect may decrease with larger turbine spacing, the reduction is unlikely to justify the additional land required. Consequently, the authors concluded that designs aimed at minimizing the cost of power generation differ from those aimed at maximizing power output per unit area.

Su et al.[24] experimentally analyzed various turbine layouts, including longitudinal and traversal groups with two turbines and a longitudinal group with three turbines. They investigated the impact of rotation on these layouts. They found that the counter-forward rotating turbine pair exhibited the best performance, surpassing other layouts with an 8% improvement over the isolated turbine at a tip speed ratio (TSR) of 1.44 and a spacing of 2.4D. In the 2-turbine longitudinal configuration, the downstream turbine's efficiency saw a significant increase of forty percent and forty-five percent at distances of 1.1D and 0.9D, respectively, compared to the isolated turbine for the counterclockwise and counter-forward rotating cases. The rotation orientation and cross-stream spacing of the preceding turbine pair were observed to have a noticeable influence on the performance of the aft turbine in the three-turbine group.

Thus, the two/three turbine cluster can be used qualitatively to understand the local wake dynamics in the wind farms. Different turbine parameters, such as turbine configurations (in-line or staggered), inter-turbine separation distances (in streamwise or transverse directions), turbine rotation, etc., affect the whole cluster's performance. Hence, in this study, the authors aim to investigate the effects of various parameters on vertical-axis wind turbines (VAWTs) using the actuator line method coupled with large-eddy simulation (LES). To achieve this, the authors customized an open-source code, Xcompact3D[17], to accurately model VAWT blades, shafts, and strut supports, which act as additional lifting surfaces. Additionally, the study considers unsteady effects like dynamic stall and added mass, significantly influencing VAWT performance compared to HAWTs.

## II. PROBLEM STATEMENT

Utilizing computationally efficient reduced order models such as ALM and ADM proves advantageous. ALM, when coupled with LES modeling, has shown accurate prediction capabilities for both steady and unsteady conditions of VAWTs, making it a valuable tool for design and optimization purposes. However, the implementation of ALM is limited to a few open-source codes. Motivated by this limitation, we have modified and validated the open-source code against available experimental and numerical data. Previous studies have highlighted the significant performance improvement achieved through turbine clustering and have investigated the effects of turbine spacing. However, limited analysis has been conducted on the impact of individual turbine rotation within a cluster on turbine efficiency. Therefore, we employed the well-validated UNH-RVAT[16] turbine to examine the influence of spacing and individual turbine rotation on performance. We have utilized the three-turbine line configuration proposed by Zhang et al.[22] and initially explored combinations of $X_{sep}$ (0D, 0.34D, 0.5D, 0.68D) and $Y_{sep}$ (4D, 3D, 2.5D) (note that Zhang et al.[22] did not report $Y_{sep}$ = 2.5D) for two three-turbines cluster configurations: staggered and inline (as shown in Fig. 1 below). The best-performing case was determined from this analysis. Subsequently, we analyzed the impact of each turbine's rotation on performance for two configurations: inline ($X_{sep}$ = 0D, $Y_{sep}$ = 2.5D) and staggered ($X_{sep}$ = 0.34D, $Y_{sep}$ = 2.5D). The primary objective of this study is to enhance performance and optimize inter-turbine spacing within turbine clusters, enabling them to operate as integrated units within wind farms and maximize energy density.

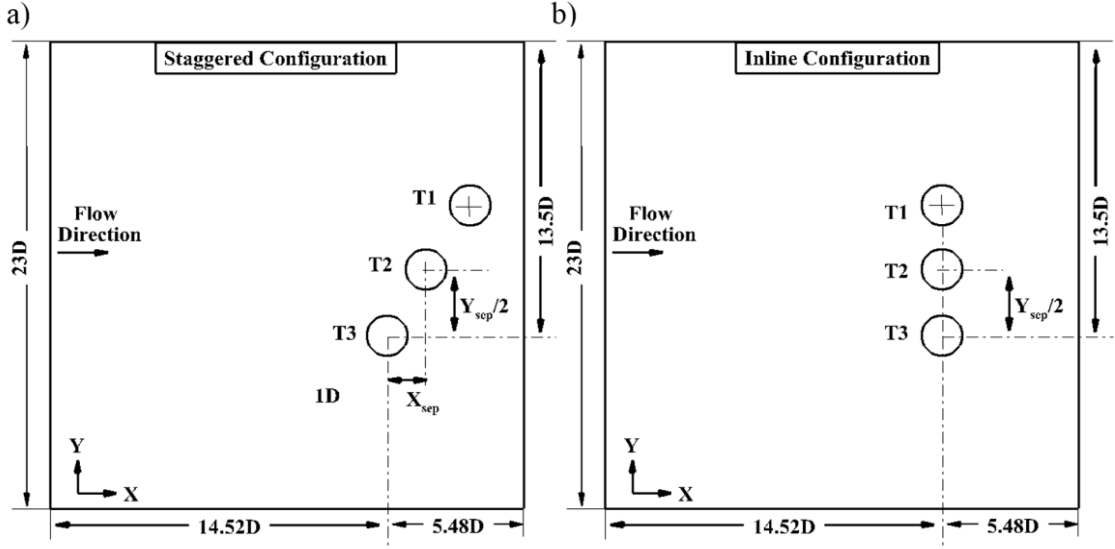

**FIG. 1.** Schematics of different configurations studied: a) staggered configuration, b) inline configuration (figures not to scale)

## III. NUMERICAL METHODOLOGY

### A. Governing Equations for Fluid

In CFD, the depiction of fluid flow involves the utilization of continuity and momentum equations. The flow is assumed incompressible, and the governing equations are provided below.

$$\frac{\partial u_i}{\partial x_i} = 0 \tag{1}$$

$$\frac{\partial u_i}{\partial t} + \frac{1}{2}\left(u_j \frac{\partial u_i}{\partial x_j} + \frac{\partial u_i u_j}{\partial x_j}\right) = -\frac{1}{\rho}\frac{\partial P}{\partial x_i} + \frac{\partial \tau_{ij}}{\partial x_j} + \frac{F_i^t}{\rho} \tag{2}$$

Where Eqs. (1) and (2) are incompressible continuity and momentum equations. Here $u_i$ and P are the velocity and pressure fields, respectively, $\rho$ is the density of the fluid, and $\tau_{ij}$ is the sub-grid stress tensor. $F_i^t$ is the turbine forcing term, which will be explained in detail in the actuator line method section. In Eqn. (2) the convective term is represented in skew-symmetric form so that the aliasing errors will get reduced.[15]

### B. Turbulence Modelling

The viscous stress tensor term in the Navier-Stokes momentum equation [Eqn. (2)] dissipate energy through molecular viscosity. It is only strong enough for the smallest (Kolmogorov) eddies. In LES, the dissipation can be increased by adding an additional stress term, i.e., the sub-grid stress[25] term $\tau_{sgs}$.

$$\tau_{ij} = \tau_{ij} + \tau_{sgs} \tag{3}$$

The sub-grid stress ($\tau_{sgs}$) is calculated using an eddy viscosity model as:

$$\tau_{sgs} = 2\rho \nu_{sgs} S_{ij}^* - \frac{2}{3}\rho \kappa_{sgs} \delta_{ij} \tag{4}$$

Where $\kappa_{sgs}$ is the sub-grid scale kinetic energy, $S_{ij}^*$ is the deviatoric stress tensor (also known as the strain rate of modeled eddies), which is defined as:

$$S_{ij}^* = \frac{1}{2}\left(\frac{\partial U_i}{\partial x_j} + \frac{\partial U_j}{\partial x_i} - \frac{1}{3}\frac{\partial U_k}{\partial x_k}\delta_{ij}\right) \tag{5}$$

The eddy viscosity ($\nu_{sgs}$) will be a scalar value considering the eddies as isotropic. Different sub-grid models use different methods to calculate the eddy viscosity. One such model is the Smagorinsky model[25] used for the present analysis. The Smagorinsky model is relatively simple and easy to implement compared to some other SGS models, such as dynamic models or more complex models like the Scale-Adaptive Simulation (SAS) model or the Wall-Adapting Local Eddy-viscosity (WALE) model. This simplicity can make it a practical choice for a wide range of LES applications. The Smagorinsky model is known for its robustness across different flow scenarios. It can provide reasonable results in various turbulent flows without extensive tuning or modification.

$$\nu_{sgs} = l_o^2 \sqrt{2 S_{ij} S_{ij}} \tag{6}$$

$$S_{ij} = \frac{1}{2}\left(\frac{\partial U_i}{\partial x_j} + \frac{\partial U_j}{\partial x_i}\right) \tag{7}$$

$$l_o = C_s \left(cellvolume\right)^{\frac{1}{3}} \tag{8}$$

Where $C_s$ is the Smagorinsky constant, and the value is always less than one so that the length scale will be smaller than the cell size, $l_o$ is the sub-grid length scale, and $S_{ij}$ is the strain rate tensor.

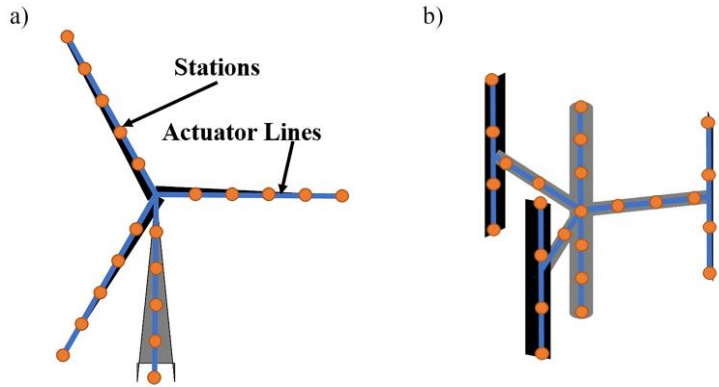

**FIG. 2.** Schematics of ALM discretization of: a) a HAWT, and b) a VAWT

## C. Actuator Line Method

Actuator line method (ALM)[16] is one of the reduced order models. It is proven that ALM on coupling with LES makes it a powerful tool to analyze wind turbine wakes and wind farms. ALM is more accurate than other reduced-order models like the Actuator Disc Model (ADM), Vortex Model, etc[27-29]. ALM considers the lifting surfaces of the turbine as lines or actuator lines. Each actuator line is divided into several stations, as shown in Figs. 2(a) and 2(b). The station is taken as an aerofoil section, and the relative velocity and the angle of attack (AOA) are calculated. The relative velocity ($\vec{U}_{rel}$) is calculated by adding the incoming flow velocity vector ($\vec{U}_{in}$) and the blade rotational velocity ($\vec{U}_b$) vector as given in Eqn. (9) and (10):

$$\vec{U}_{rel} = \vec{U}_{in} + \vec{U}_b \qquad (9)$$

$$\vec{U}_b = \omega \vec{r} \qquad (10)$$

where $\vec{r}$ is the radius of the rotor, and ω is the rotor's angular velocity. On calculating the relative velocity, Reynolds's number (chord-based) is known now in that section. The static coefficients (lift, drag, and moment coefficients) are then interpolated from the data table with the coefficients for a range of Reynolds's number and AOA. The static coefficients utilized in this study are sourced from the Sheldahl and Kilmas database[30], which offers a comprehensive collection of values across a wide range of Reynolds numbers. The dynamics will not be static as the rotor blade continuously rotates in the fluid. So, the unsteady nature must be added and modeled to represent the actual flow conditions. The unsteady effects like a dynamic stall, flow curvature, added mass, and end effects have been added and discussed in the coming sub-sections.

## D. Dynamic Stall

Dynamic stall occurs when an aerofoil changes its angle of attack with time (i.e., unsteady pitching). As the rotor blade sections continuously change their AOA, they are subjected to dynamic stall. When the aerofoil pitches continuously, a trailing edge vortex will be formed. This vortex induces a flow reversal at the trailing edge in the boundary layer. This, in turn, induces Kelvin Helmholtz instability; as a result, a leading-edge vortex strengthens. A leading-edge vortex traversing the upper surface of the aerofoil enhances the lift coefficient beyond the static stall lift coefficient. The nose pitch-down moment coefficient drastically increases. These increased loads beyond the static stall induce additional torque on the rotor. The leading-edge vortex gets separated, leaving the aerofoil stalled, and the stall angle is beyond the static stall angle. So, the whole process involves three processes: attached flow, vortex-induced flow, and separated flow, as discussed extensively in Leishman's work[31]. Leishman and Beddoes[32] developed an actuator line method-compatible dynamic stall model that has found widespread usage in the aerodynamics of helicopters and wind turbines. They fine-tuned it for helicopter applications with Mach numbers greater than 0.3. However, analyses based on experiments have shown that the model provides less accurate reconstructions of the unsteady air loads at low Mach numbers. Sheng et al.[33] altered the model to be effective for low Mach numbers and validated extensively with the experimental results employed in the present analysis.

**E. Flow Curvature**

Since the rotor is in orbital motion, the blade sections' angle of attack over the chord will not be constant. Defining a singular AOA for accessing the static tables in this circumstance is problematic. Goude et al.[34] proposed a model for flow curvature by contemplating a circularly moving flat plate in potential flow. The modified angle of attack after considering flow curvature effects is as given in Eqn. (11) below:

$$\alpha = \delta + \arctan\left(\frac{V_{in} \cos(\theta - \beta)}{V_{in} \sin(\theta - \beta) + \omega r}\right) - \frac{\omega x_c c}{V_{ref}} - \frac{\omega c}{4V_{ref}} \tag{11}$$

The variables mentioned include pitch angle ($\delta$), incoming velocity magnitude ($V_{in}$), azimuthal angle ($\theta$), incoming velocity direction ($\beta$), rotor angular velocity ($\omega$), blade attachment point along the chord ($x_c$), chord length ($c$), and free stream velocity ($V_{ref}$). As the blade section AOA is determined using vector operations, the first two elements of the Eqn. (11) are implicitly included. It is, therefore, sufficient to add the last two terms to the calculated AOA.

**F. Added Mass**

When an immersed object accelerates through a fluid, there will be a net displacement of fluid. As a result of this displacement, the fluid exerts an inertia force on the submerged object. The added mass effects become apparent as the VAWT's blades and struts accelerate through the fluid. Strickland et al.[35] proposed a method for calculating these supplementary mass effects. They viewed the blades as flat plates and calculated the coefficients based on the rotation of these plates in a potential flow. The normal force, tangential force, and moment due to added mass are given as follows:

$$F_{CAM} = \pi \rho \frac{c^2}{4} \omega U_n \tag{12}$$

$$F_{NAM} = -\pi \rho \frac{c^2}{4} \frac{dU_n}{dt} \tag{13}$$

$$M_{AM} = -\pi\rho U_n U_c \frac{c^2}{4} \tag{14}$$

Where ρ is the density of the fluid, $F_{CAM}$ is the tangential force, $F_{NAM}$ is the normal force, and $M_{AM}$ is the moment about a mid-chord, $U_n$ and $U_c$ are the normal and tangential components of resultant velocity, respectively. The moment at the quarter chord point ($M_{c/4}$) is given by Eqn. (15):

$$M_{c/4} = \frac{c}{4} F_{NAM} - \pi\rho U_n U_c \frac{c^2}{4} \tag{15}$$

The normal ($C_{NAM}$), tangential ($C_{CAM}$), and moment ($C_{MAM}$) coefficients are then calculated from the forces and moments [Eqs. (12)-(15)]. They are then converted into the lift and drag coefficients and added to the coefficients obtained from the dynamic stall ($C_L, C_D, C_M$).

$$C_{lAM} = C_{NAM} \cos\alpha + C_{CAM} \sin\alpha \tag{16}$$

$$C_{dAM} = C_{NAM} \sin\alpha - C_{CAM} \cos\alpha \tag{17}$$

$$C_L = C_L + C_{lAM} \tag{18}$$

$$C_D = C_D + C_{dAM} \tag{19}$$

$$C_M = C_M + C_{MAM} \tag{19}$$

### G. End Effects

According to the Kelvin-Helmholtz vortex theorem, a vortex line should either form a closed loop or extend to the boundaries of a fluid. Therefore, the lift distribution resulting from a bound vortex should be zero at the ends. Since the coefficients are interpolated from the static data, the lift at the extremities of the blades is not zero. Bachant et al.[16] have proposed an end effects model, which is relatively simple compared to the famous Glauert's model[36]. Bachant's model takes into account the lifting line theory proposed by Prandtl. Based on this theory, the aerofoil AOA with any circulation distribution can be expressed as a function of the normalized span $\theta$.

$$\alpha(\theta) = \frac{2S}{\pi c} \sum_{n=1}^{N} A_n \sin\theta + \sum_{n=1}^{N} nA_n \frac{\sin n\theta}{\sin\theta} + \alpha_{L=0} \tag{20}$$

Where $S$ is the span length, and $N$ is the number of stations. The Fourier coefficients are the only unknown ($A_n$) and can be found by solving the matrix equations. The circulation distribution can be found using the following equation.

$$\Gamma(\theta) = 2SU_\infty \sum_{n=1}^{N} A_n \sin n\theta \qquad (21)$$

Once the circulation is known, the lift coefficient can be evaluated from the Kutta-Joukowski theorem.

$$C_l = \frac{-\Gamma(\theta)}{\frac{1}{2}cU_\infty} \qquad (22)$$

The end effect factors (F) can be found by normalizing the lift coefficient obtained at each aerofoil section with the max lift coefficient obtained.

$$F = \frac{C_l}{C_{l_{max}}} \qquad (23)$$

Before calculating the forces operating on the blade section, the lift coefficient is scaled using the end effect factor. Glauert's model considers the blade section AOA and tip speed ratio (TSR) when calculating the end effects, whereas Bachant's model does not.

### H. Power Coefficient

The forces acting on each blade in all three directions are calculated, and the torque produced $\vec{\tau}$ is evaluated in the equation below.

$$\vec{\tau} = \vec{r} \times \vec{F} \qquad (24)$$

Where $\vec{r}$ is the radius vector from the center of the turbine to the blade section, and $\vec{F}$ is the force vector acting on that blade section.

$$P = \vec{\tau} \cdot \vec{\omega} \qquad (25)$$

Where $P$ is the power produced and $\vec{\omega}$ is the angular velocity of the turbine. The power coefficient is evaluated by dividing the power produced with the available power, as shown below.

$$C_p = \frac{P}{0.5 \rho A U_\infty^3} \qquad (26)$$

Where $\rho$ is the density of the fluid, $A$ is the frontal area of the turbine and $U_\infty$ is the free stream velocity of the fluid. Further, the tip speed ratio can be defined as:

$$TSR = \frac{r\omega}{U_\infty} \tag{27}$$

**I. Force as Source**

The forces acting on each blade section, when evaluated, are then incorporated into the momentum equation [Eqn. (2)] as a source term. The force is spread out over a region using the spherical Gaussian function, which the Eqn. (28) provides. This ensures that strong gradients do not lead to instability when the force term is included.

$$\eta_{ij} = \frac{1}{\varepsilon^3 \pi^{\frac{3}{2}}} e^{\left(-\left(\frac{\eta_{ij}}{\varepsilon}\right)^2\right)} \tag{28}$$

where $\eta_{ij}$ is the Gaussian width, $\varepsilon$ is a smoothing parameter, and $r_{ij}$ is the distance from the mesh node to the actuator line node. The extent of the Gaussian function is regulated by the filtering parameter $\varepsilon$, as depicted in Eqn. (29). Martinez et al.[37] conducted a study in which they identified the chord length, drag force, and grid size as the key factors to be considered while determining the appropriate smoothing parameter. Troldborg[38] suggested that the Gaussian width should be twice the size of the local mesh to ensure numerical stability. Using the cell size that contains the element, we can calculate the $\varepsilon$ value defined by the mesh size as shown below:

$$\varepsilon = 2C_{mesh}\Delta \tag{29}$$

$$\Delta = \sqrt[3]{\Delta x \Delta y \Delta z} \tag{30}$$

After reviewing the literature[16], $C_{mesh}$ is assumed to be 2.0 for the simulations. The drag force's Gaussian width is proportional to the momentum thickness $\theta$.

$$\theta = \frac{C_d c}{2.0} \tag{31}$$

where $C_d$ is the drag coefficient. The maximum chord length, mesh size, and drag force are used to calculate the Gaussian width for force projection. This ensures that coarse meshes will not cause instability, while fine mesh improves precision. The flow chart of the whole ALM is given in the appendix.

**IV. RESULTS AND DISCUSSION**

The present study is divided majorly into five subsections. The first subsection extensively validates the modified solver for VAWTs calculations by comparing it with available experimental and numerical data. The second subsection deals with the mesh independence study and error analysis. In the third subsection, the author explores the effect of the computational domain blockage ratios on the performance assessment of the VAWTs cluster by varying the boundaries of the computational domains in streamwise and transverse directions. Further, the authors perform the simulations for varying the inter-turbine distances corresponding to two configurations (i.e., inline and staggered) to optimize the turbine cluster's power density. In the final subsection (i.e., subsection IV. E), the authors vary the rotation direction of the turbines and assess its effect by examining the pressure contours, velocity plots, and the kinetic energy deficit across the turbines by fixing the $X_{sep}$ and $Y_{sep}$ distance between the turbines (for both inline and staggered configurations).

## A. Code Validation and Verification for the Single Turbine

This subsection is devoted to validating the modifications incorporated into the available solver with the available numerical and computational literature. The numerical simulations have been performed over the computational domain of dimensions $L_x = 3.68D$, $L_y = 3.66D$, and $L_z = 2.44D$ (where D is the diameter of the rotor) and shown in Fig. 3 (a) and (b) along with the turbine details, which corresponds to the actual experimental test section reported by Bachant et al.[16]. The turbine parameters are given in Table I. The number of elements per actuator line was set to be approximately equal to the total span divided by the Gaussian force projection width $\varepsilon$ employed by Bachant et al. [16].

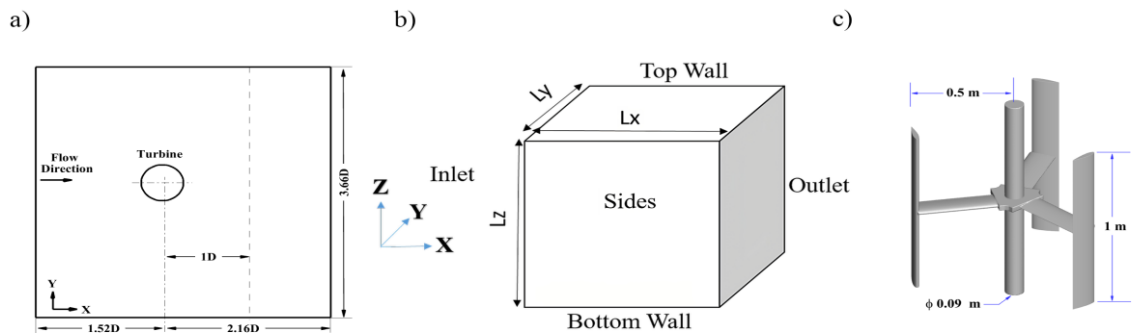

**FIG. 3.** Schematics of the computational domain used for the validation study: a) x-y plane for demarcation of different dimensions, b) Isometric view of the UNH-RVAT Turbine

We have adopted the same boundary conditions as Bachant et al.[16] (see Table II). The experimental setup of Bechant et al.[16] involves towing the turbine through a water tunnel. As a result, the top boundary is open to the atmosphere while the side and bottom boundaries are moving at a speed of 1 m/s. According to the literature [16,42-44], the suggested grid size for ALM combined with LES is 18–64 cells per turbine rotor diameter. We have thus considered a grid resolution of 64 cells per rotor diameter in all directions, resulting in a total grid size of approximately 8.6 million. The turbine center is the origin, and the inlet and outlet are 1.52D upstream and 2.16D downstream of the rotor. The turbine is placed in the center along the spanwise direction.

**TABLE I.** Parameters of UNH-RVAT Turbine

| Parameter | Value |
|---|---|
| Rotor diameter (D) | 1 m |
| Blade span | 1D |
| Rotor solidity | 0.28 |
| Blade chord (c) | 0.14D |
| Shaft diameter | 0.09D |
| Number of blades | 3 |
| Pitch angle of blade sections | $0°$ |
| TSR ($\lambda$) | 1.9 |
| Airfoil | NACA0021 |

**TABLE II.** Boundary conditions used for the validation study

| Boundary | Boundary Condition | Value |
|---|---|---|
| Inlet | Velocity inlet | 1 m/s |
| Outlet | Convective outflow | - |
| Sides | Moving wall | 1 m/s |
| Top wall | Free slip | - |
| Bottom wall | Moving wall | 1 m/s |
| Kinematic viscosity | - | $1e^{-6}$ |

The simulations have been run for 10 seconds with a time-step of 0.002 s. The second-order Adam-Bash forth time integration scheme is used, and sixth-order compact schemes are used to discretize the convection and diffusion terms. The statistics are averaged after 5 seconds of simulation time to exclude the initial transient fluctuations. The rotor's tip speed ratio is adjusted to oscillate with a magnitude of 0.19 and a phase angle of 0.14 radians about the mean tip speed ratio ($\lambda_0$ = 1.9).

Fig. 4 (a) shows the mean velocity ($U_{mean} = \frac{1}{T}\int_0^T U_x dt$, where T is the time period for the averaging and $U_x$ is the streamwise component of the instantaneous velocity) profile in the turbine's wake at a downstream location of 1D from the rotor center where the unsteady effects (dynamic stall, added mass, end effects, flow curvature) are not considered. The wake effects are poorly recorded compared to the experimental findings of Bachant et al.[38]. This is attributed to the interpolations made using static 2-D data and the consideration of the blade sections as airfoils. This will not consider the three-dimensional effects critical in determining the rotor's wake expansion. Upon adding dynamic stall effects, there is a quite improvement in the wake profile, as shown in Fig. 4 (b). The normalized wake velocity profile at a downstream distance of 1D from the rotor with all the unsteady effects invoked is shown in Fig. 5 (a). It is observed that the wake characteristics are well determined and are in good agreement with the experimental and numerical findings of Bachant et al.[16]. The error is lesser in the present case than in the case of Hezaveh et al.[18].

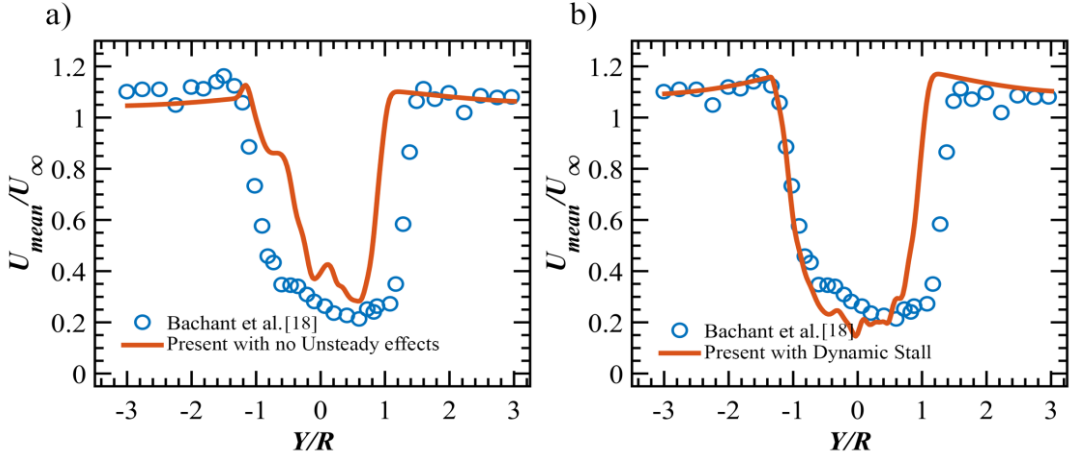

**FIG. 4.** Normalized mean velocity in streamwise direction ($U_{mean}$) normalized with free stream velocity ($U_\infty$) profile at x/D=1 downstream of turbine: a) without Unsteady Effects, and b) with dynamic stall invoked

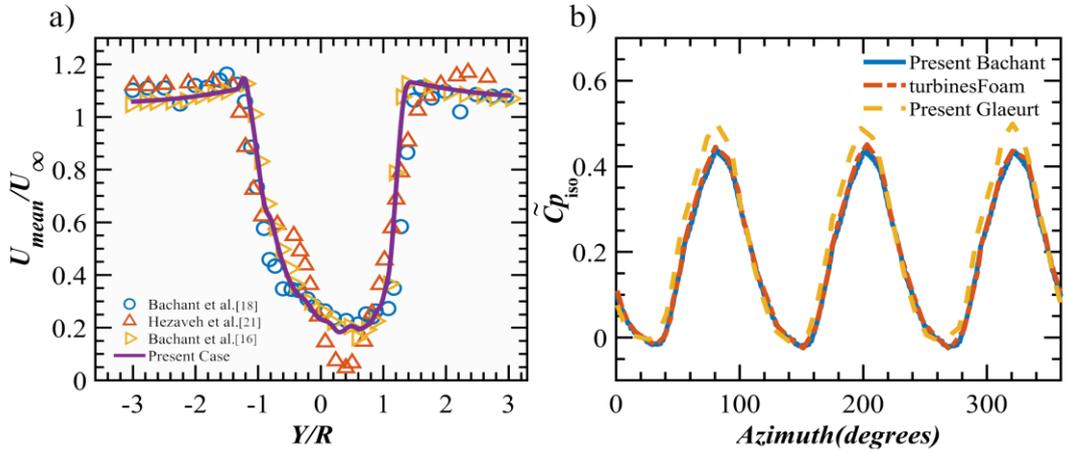

**FIG. 5.** Comparison study with the available literature for: a) mean velocity in streamwise direction ($U_{mean}$) normalized with free stream velocity ($U_\infty$) profile at x/D=1 downstream of the turbine, and b) Instantaneous power coefficient ($\tilde{C}_{p_{iso}}$) variation over a cycle.

Bachant et al.'s[16] and Glauert's[36] end effects models, both extensively discussed in the numerical methodology section, are the two models considered for the current validation. The rotor's instantaneous power coefficient ($\tilde{C}_{p_{iso}}$) throughout a rotation cycle is displayed in Fig. 5 (b) alongside the turbinesFoam[16] findings. The power coefficient calculated using Bachant's end effect model[16] is close to that calculated using turbinesFoam[16]. This demonstrates that the solver modifications are in excellent accord. The experimental mean power coefficient of 0.25 is more comparable to the mean power coefficient estimated using Glauert's end effect model[36], which is 0.21. This is because Glauert's model[36], which is more robust and accurate, accounts for the rotor's blade section AOA and TSR when computing the end effects. Based on the above discussions, the authors have adopted Glauert's end-effect model for further simulations.

**B. Mesh Independence and Error Analysis**

Before continuing to deal with the turbine spacing and rotation analysis, we have performed the grid independence study for the three-turbine cluster configuration. The present study deals with two different turbine-cluster configurations: staggered and inline (as shown in Fig. 1 (a) and (b)). The mesh independence study considers the staggered configuration of $X_{sep}$ = 0.34D and $Y_{sep}$ = 2.5D with the computational domain dimensions shown in Fig. 1 (a) and the boundary conditions tabulated in Table III. The information related to the turbine's (i.e., UNH-RVT) parameters is listed in Table I.

**Table III.** Boundary Conditions considered for the simulations

| Boundary | Boundary Conditions | Values |
|---|---|---|
| Inlet | Velocity inlet | 1 m/s |
| Outlet | Convective outflow | - |
| Walls | Free slip | - |
| Kinematic Viscosity | - | $1e^{-6}$ |

The simulations are performed over four different grids: Grid-A (with ~39 million cells), Grid-B (with ~76 million cells), Grid-C (with ~148 million cells), and Grid-D (with ~290 million cells). The mean power coefficient ($C_p$) for the three-turbine cluster is compared for different grids in Table IV and in Fig. 6. We can see that the difference between grid-C and grid-D is less than 2%; hence, grid-C (with 148 million cells) seems good enough to carry out the simulations.

**TABLE IV.** Grid independence study for the three-turbine cluster (values in bold show the selected grid)

| Grid | No. of Cells | $C_p$ | % change in $C_p$ value |
|---|---|---|---|
| Grid-A | 39,317,017 | 0.192 | - |
| Grid-B | 76,150,269 | 0.184 | 4.347 % |
| **Grid-C** | **148,885,737** | **0.178** | **3.370 %** |
| Grid-D | 290,233,251 | 0.176 | 1.136 % |

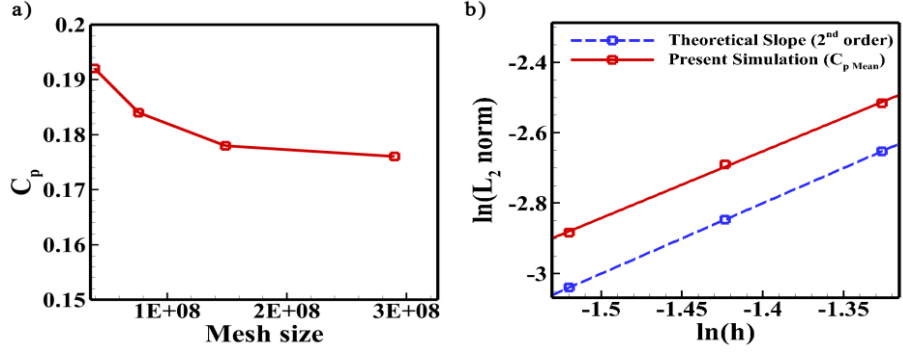

**FIG. 6.** Grid independence study for four different mesh sizes: a) variation of the mean power coefficient with the mesh size, b) L$_2$-norm against the grid spacing parameter h, compared to the theoretical slope of 2$^{nd}$ order

To gain more confidence in the grid sizing, the Grid-Convergence Index (GCI) calculations are performed by considering the mean power coefficient ($C_p$) as the parameter, proposed by Roache[39, 40], and has been extensively used in the past literature. GCI estimates the error based on the generalized Richardson extrapolation-derived grid refinement. The Grid Convergence Index (GCI) indicates the potential change in the computed value with further refinement. A lower GCI percentage signifies that the computed value converges toward the asymptotic range. We have considered grid-B, grid-C, and grid-D for the GCI calculations, and the results are tabulated in Table II. We find the GCI for the fine grid is less than 5%, confirming that the grid-C is nicely resolved to perform these simulations. The detailed formulations of the GCI can be found in the previous work of the research group[41].

**TABLE V.** Richardson error estimation and GCI calculations for the three sets of grids

|  | $r_{CB}$ | $r_{DC}$ | o | $\varepsilon_{CB}(\times 10^{-2})$ | $\varepsilon_{DC}(\times 10^{-2})$ | $E_2^{coarse}$ | $E_1^{fine}$ | $GCI^{coarse}$ | $GCI^{fine}$ |
|---|---|---|---|---|---|---|---|---|---|
| $C_p$ | 1.25 | 1.25 | 1.897 | -3.370 | -1.136 | 0.097 | 0.021 | 12.125 % | 2.625 % |

**C. Effect of the Computational Domain Boundaries for the Three-turbine Cluster**

Before continuing with the detailed study, we have conducted a domain optimization study to ensure that boundary effects of the computational domain don't affect the simulation results. The boundary conditions for this domain optimization study are shown in Table III, keeping the turbine parameters the same as in Table I. The turbine-cluster configuration used in this study is staggered (X$_{sep}$ = 0.34D and Y$_{sep}$ = 2.5D), as shown in Fig. 1 (a). We first looked at the effect of the cross-stream boundaries of the computational domain on the turbine cluster's performance calculations. Fig. 7 (a) reports the variation of the power coefficient of the cluster for two distinct domain lengths in the cross-stream direction (L$_y$ = 23D and 33D) over a range of streamwise separations (X$_{sep}$) values. The observed performance difference from 23D to 33D is less than 0.05% over the entire range, indicating the negligible effect of the cross-stream boundaries of the computational domain. Therefore, we have considered a 23D domain length in the cross-stream direction for further simulations.

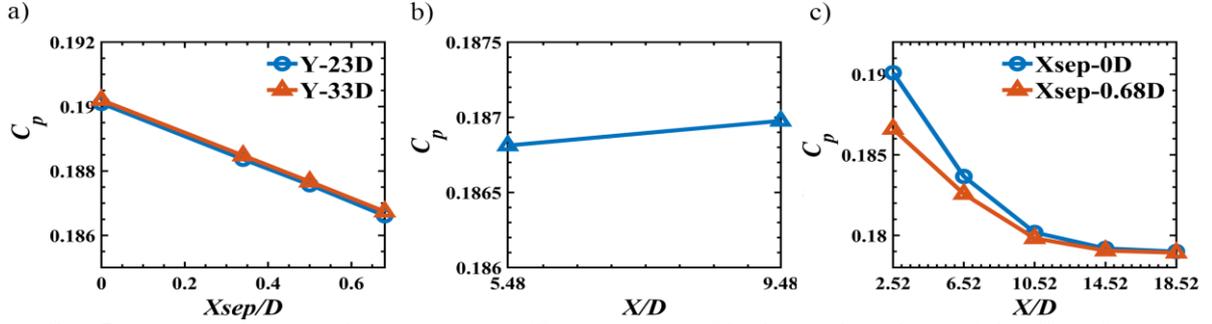

**FIG. 7.** Domain optimization study using power coefficient variation of the three-turbine cluster: a) the effect of cross-stream boundaries, b) the effect of the placement of outlet boundary from the center of turbine T3, and c) the effect of the inlet boundary distance from the turbine rotor

Further, we have examined the effect of the outlet boundary for the case of $X_{sep}$ = 0.68D and cross-stream separation ($Y_{sep}$) = 4D by altering the downstream distance of the outlet boundary from the rotor in Fig. 7 (b). The performance variation for the downstream domain lengths of 5.48D and 9.48D is less than 0.03 percent. Hence, we have chosen a length of 5.48D for further simulations. The performance variation across two streamwise spacings (0D and 0.68D) for various inlet distances (i.e., the distance of the inlet from the center of the rotor) is shown in Fig. 7 (c). The intake impacts performance up to 14.52D, beyond which there is barely any change in performance. Therefore, the upstream inlet distance of 14.52D is considered for subsequent study. We have considered the $Y_{sep}$ of 4D for each situation described above since this is the largest cross-stream separation examined. The domain size along the spanwise direction of the turbine is considered to be 2.5D.

**D. Optimizing the Spacing Between Turbines**

We have utilized the optimized domain to examine how different streamwise ($X_{sep}$ = 0D, 0.34D, 0.5D, 0.68D) and transverse spacings ($Y_{sep}$ = 2.5D, 3D, 4D) between the individual turbines affect the performance of the turbine clusters. We specifically reduced the transverse spacing to 2.5D, whereas Zhang et al.[22] only reported data up to 3D. Extensive analysis is conducted on various streamwise and transverse spacing combinations to determine the most favorable configuration from the improved performance perspective. The boundary conditions used in this study are identical to those listed in Table III. For this analysis, all the turbines are considered to be co-rotating in a counter-clockwise direction.

We first evaluated the performance of a single turbine by simulating it under conditions similar to the turbine clusters. The achieved power coefficient for the isolated turbine is 0.161. The variations in the power coefficient ratio ($C_p / C_{p_{iso}}$), which represents the ratio of the mean power coefficient of the turbine cluster to the power coefficient of an isolated turbine, are illustrated in Fig. 8 (a) for different combinations of streamwise ($X_{sep}$) and transverse ($Y_{sep}$) separations. A significant performance improvement, reaching a minimum of 8.2%, is observed for the case of $X_{sep} = 0.68D$ and $Y_{sep} = 4D$. The power coefficient shows an increasing trend as the streamwise separation decreases until it reaches 0.34D. Beyond this point, the power coefficient either decreases or remains constant as the separation decreases.

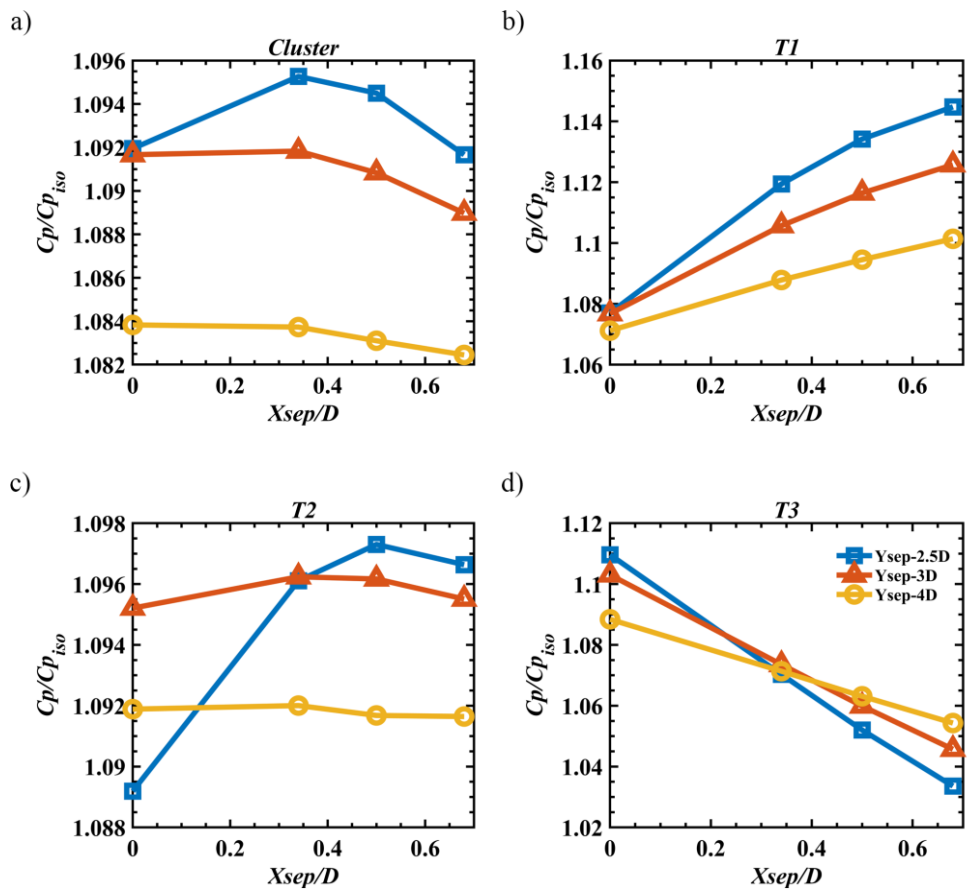

**FIG. 8.** Normalized mean power coefficients of: a) a three-turbine cluster rotating in counter clockwise direction, b) Turbine T1, c) Turbine T2, and d) Turbine T3.

Furthermore, the power coefficient increases with lower $Y_{sep}$ values. This phenomenon can be explained by the increased flow acceleration between the turbines as the spacing reduces, resulting in improved performance. The largest performance improvement is recorded in the cases of $X_{sep} = 0.34D$ and $Y_{sep} = 2.5D$, with a 9.52% improvement compared to the isolated turbine. For cross-stream separations of 3D and 4D, the performance gradually improves as the streamwise separation ($X_{sep}$) decreases until 0.34D, which remains constant as the separation decreases further to 0D. However, for $Y_{sep}$ = 2.5D, the performance gradually increases from $X_{sep}$ = 0.68D to 0.34D but decreases as the streamwise separation distance decreases to 0D.

In this specific line configuration, T1 is positioned as the downwind turbine, T3 as the upwind turbine, and T2 as an intermediate turbine between T3 and T1. The performance variation of the downstream turbine T1 is illustrated in Fig. 8 (b). Across all $Y_{sep}$ scenarios, there is an improvement in performance as $X_{sep}$ increases. This phenomenon can be attributed to the turbine leveraging the enlarged wake as it progresses downstream from turbine T2, resulting in enhanced performance.

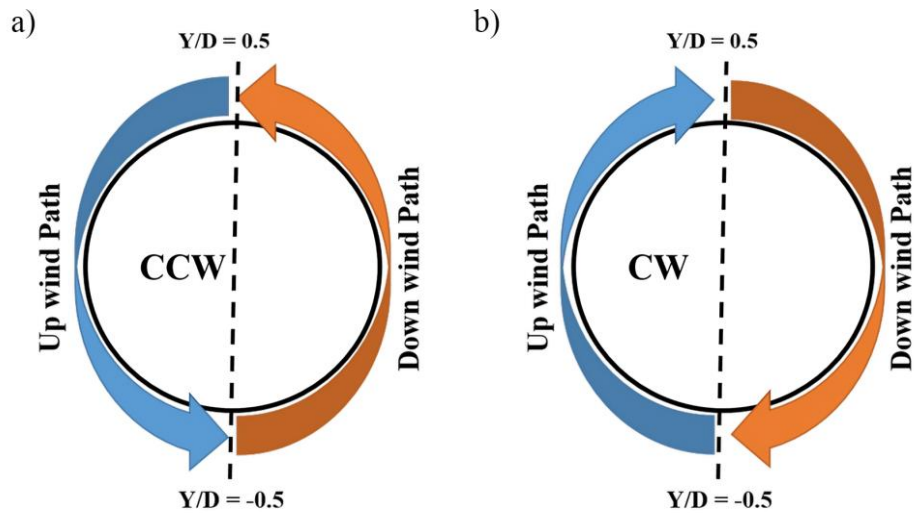

**FIG. 9.** Schematics of upwind and downwind paths for: a) Counter Clockwise Rotating turbine and b) Clockwise Rotating turbine

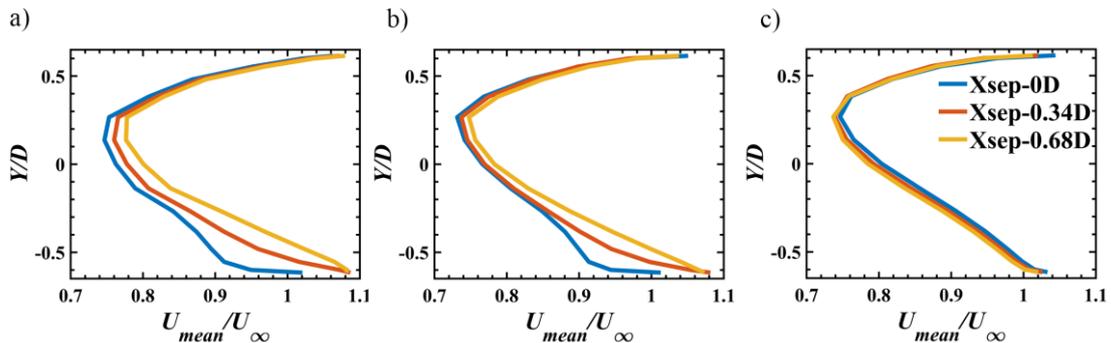

**FIG. 10.** Normalized stream-wise mean velocity (for $Y_{sep} = 2.5D$) over the upwind side of: a) Turbine T1, b) Turbine T2, and c) Turbine T3

Fig. 9 (a) and (b) show the schematic of the upwind and downwind paths for counterclockwise and clockwise (CCW and CW) rotating turbines. The expansion of the wake causes an increase in the available kinetic energy, enabling the turbine to capture more energy from the flow. Fig. 10 illustrates the variation in mean velocity on the upwind side of the turbines. It is observed that the velocity increases on the left half of the upwind side, indicating that the expansion of the wake leads to an increase in available energy for the downstream turbine T1. Additionally, from Fig. 11 (a), it can be observed that the kinetic energy downstream of the turbine decreases as the $X_{sep}$ increases. This decrease indicates that the energy extraction from the flow is increasing.

This trend is further supported by Fig. 12 (a), demonstrating the kinetic energy change across the turbine. The kinetic energy is calculated by spatially averaging over a line 1D upstream (Kinetic energy available) and 1D downstream (Kinetic energy unused) of the turbine for every time step, and the formulation is as given in eq. (32). The performance of turbine T1 improves as the cross-stream separation decreases. This improvement can be attributed to the reduced distance between the turbines, allowing the turbine to interact closely with the bypass region, which possesses higher energy than the free stream flow.

$$KE = 0.5U^2 \tag{32}$$

Where $KE$ is the kinetic energy and $U$ is the spatially averaged velocity.

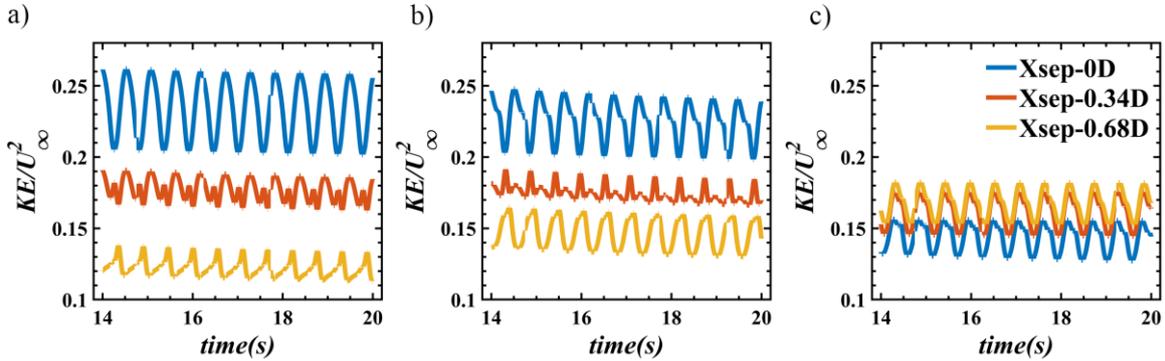

**FIG. 11.** Normalized kinetic energy (for $Y_{sep} = 2.5D$) at 1D downstream: a) Turbine T1, b) Turbine T2, and c) Turbine T3.

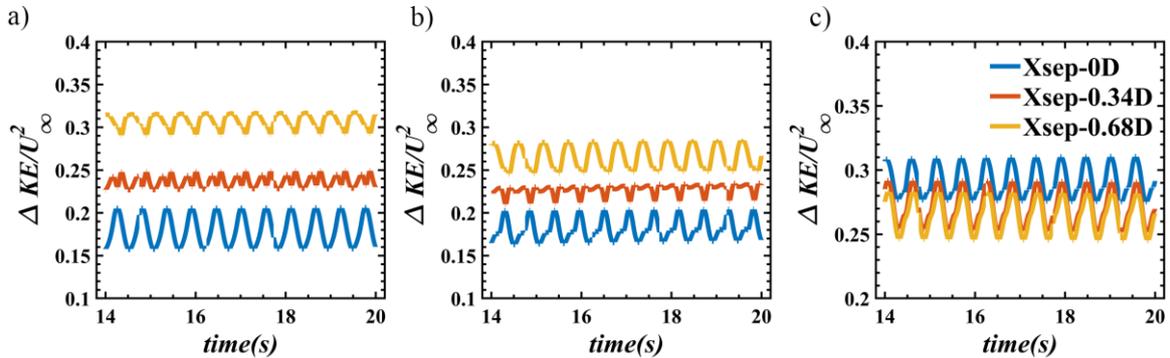

**FIG. 12.** Normalized kinetic energy change ($\Delta KE$ for $Y_{sep} = 2.5D$) over: a) Turbine T1, b) Turbine T2, and c) Turbine T3

The performance of the upstream turbine, T3, is noticeably different from that of the downstream turbine, T1, as shown in Fig. 8(d), where a continuous decrease in performance is observed on increasing $X_{sep}$ for all $Y_{sep}$ values. This disparity arises because T3 consistently faces the same flow conditions, resulting in a lack of variation in available kinetic energy. This is demonstrated in Fig. 10 (c), where the velocity remains constant across all $X_{sep}$ cases. However, the presence of the intermediate turbine, T2, influences the expansion characteristics of the wake as the stream-wise separation changes. T2 creates a counteractive effect that restricts the complete expansion of the wake. This counteractive effect becomes more pronounced with increasing cross-stream separation, as depicted in Fig. 11 (c), which illustrates that the kinetic energy at a distance of 1D downstream of the turbine increases with larger stream-wise separation. As a result of this counteractive effect, the available kinetic energy cannot be fully harnessed by the turbine. Fig.12 (c) displays the variations in kinetic energy across turbine T3, showing a decreasing trend as the stream-wise separation increases. Consequently, reduced energy extraction from the flow leads to a decrease in net power output.

The performance of Turbine T2, positioned between Turbine T1 and T3, exhibits a distinctive pattern that differs from the other two turbines, with a consistent trend of either increasing or decreasing performance. Upon analyzing Fig. 8 (c), it becomes evident that when there is a $Y_{sep}$ = 2.5D and $X_{sep}$ = 0.34D, there is a significant and noticeable improvement in performance compared to the other transverse and streamwise separations. This case of $X_{sep}$ = 0.34D and $Y_{sep}$ = 2.5D stands out among all other cases due to its remarkable performance enhancement. In contrast, for all the other $Y_{sep}$ cases, the performance does not change significantly when altering $X_{sep}$, resulting in an overall decrease in performance for those cases with increasing $X_{sep}$. The range of performance variation observed in Turbine T2 aligns with the overall performance of the entire turbine cluster, indicating a similar behavior. Therefore, it can be concluded that the intermediate turbine, T2, plays a crucial role in determining the cluster's overall performance.

Through our analysis, we have discovered that the one with a counterclockwise rotation of turbines and a separation distance of $X_{sep}$ = 0.34D and $Y_{sep}$ = 2.5D yielded the best performance among the various cases studied. This finding is significant because previous research by Zhang et al.[22] did not report a $Y_{sep}$ value less than 3D as favorable. The performance improvement observed when transitioning from $Y_{sep}$ = 3D to $Y_{sep}$ = 2.5D was moderate. However, the reduction in turbine spacing by 0.25D resulted in a notable increase in power density, which aligns with the primary objective of our research. The combined effect of enhanced performance and increased power density is substantial and holds considerable potential. These findings highlight the significance of optimizing turbine spacing for maximizing energy production in wind farms. By utilizing a counterclockwise turbine rotation and carefully selecting the $X_{sep}$ and $Y_{sep}$ values, we can achieve improved performance and power density, leading to more efficient and productive wind energy generation.

**E. Effect of Turbine Rotation Direction**

This section presents the results obtained from the analysis of variations in the rotation directions of the turbines. In the cluster, the turbines are individually rotated in clockwise (C) and counterclockwise (CC) directions. The analysis extensively examines the combinations of these rotations for two distinct cases: inline configuration (with a separation distance of $X_{sep} = 0D$ and $Y_{sep} = 2.5D$) and staggered configuration (with a separation distance of $X_{sep} = 0.34D$ and $Y_{sep} = 2.5D$). This section involves studying the performance and behavior of the turbines under the different configurations tabulated in Table VI [where, CC: Counter Clockwise rotation, C: Clockwise rotation].

**Table VI.** Combinations of rotations of three different turbines

| Case | Turbine T1 | Turbine T2 | Turbine T3 |
|---|---|---|---|
| CC-CC-CC | CC | CC | CC |
| C-C-C | C | C | C |
| C-CC-C | C | CC | C |
| CC-C-CC | CC | C | C |
| CC-CC-C | CC | CC | C |
| CC-C-C | CC | C | C |
| C-CC-CC | C | CC | CC |
| C-C-CC | C | C | CC |

The rotation effects in both directions are thoroughly examined for the inline configuration, where the turbines are arranged in a straight line. The aim is to understand how the counter-rotating turbines influence each other and whether the arrangement enhances or hinders the overall performance. Similarly, for the staggered configuration, where the turbines are offset from each other, the counter-rotating analysis explores the interactions between the turbines and the impact of rotating them in opposite directions. Overall, this analysis of counter-rotating turbines provides valuable insights into the optimal configurations and rotation directions for maximizing the performance and efficiency of wind turbine systems. The different combinations of these three turbine rotations resulted in eight cases, as shown in Table VI.

**Table VII.** Combinations of rotations of three different turbines for staggered configuration

| Case | $C_p / C_{p_{iso}}$ | | | |
|---|---|---|---|---|
| | T1 | T2 | T3 | Cluster |
| CC-CC-CC | 1.119 | 1.096 | 1.073 | 1.096 |
| C-C-C | 1.145 | 1.087 | 1.033 | 1.088 |
| C-CC-C | 1.175 | 1.04 | 1.053 | 1.089 |
| CC-C-CC | 1.112 | 1.138 | 1.024 | 1.091 |
| CC-CC-C | 1.12 | 1.097 | 1.066 | 1.094 |
| CC-C-C | 1.115 | 1.113 | 1.043 | 1.09 |
| C-CC-CC | 1.174 | 1.039 | 1.057 | 1.09 |
| C-C-CC | 1.143 | 1.112 | 1.011 | 1.089 |

**Table VIII.** Combinations of rotations of three different turbines for inline configuration

| Case | $C_p / C_{p_{iso}}$ | | | |
|---|---|---|---|---|
| | T1 | T2 | T3 | Cluster |
| CC-CC-CC | 1.078 | 1.089 | 1.111 | 1.093 |
| C-C-C | 1.111 | 1.089 | 1.078 | 1.093 |
| C-CC-C | 1.117 | 1.101 | 1.088 | 1.102 |
| CC-C-CC | 1.088 | 1.101 | 1.117 | 1.102 |
| CC-CC-C | 1.083 | 1.107 | 1.095 | 1.095 |
| CC-C-C | 1.095 | 1.107 | 1.083 | 1.095 |
| C-CC-CC | 1.113 | 1.082 | 1.105 | 1.10 |
| C-C-CC | 1.105 | 1.082 | 1.113 | 1.10 |

Tables VII and VIII show the individual turbine power coefficient ratios for the different cases in staggered and inline configurations. The results reveal a decrease in performance within the staggered configuration, even when considering the co-rotating case (C-C-C, where all turbines rotate in the clockwise direction) compared to the co-rotating anti-clockwise rotating case (CC-CC-CC). Fig. 14 (a) illustrates the pressure distribution over the upwind sides of the three turbines in the staggered anti-clockwise co-rotating [CC-CC-CC], while Fig. 14 (b) depicts the pressure distribution in the staggered clockwise co-rotating clockwise case [C-C-C]. The contour plots reveal that in the C-C-C case, the high-pressure regions tend to form near the areas where the flow accelerates between the turbines. However, these regions are relatively farther from the flow acceleration zones in the CC-CC-CC case.

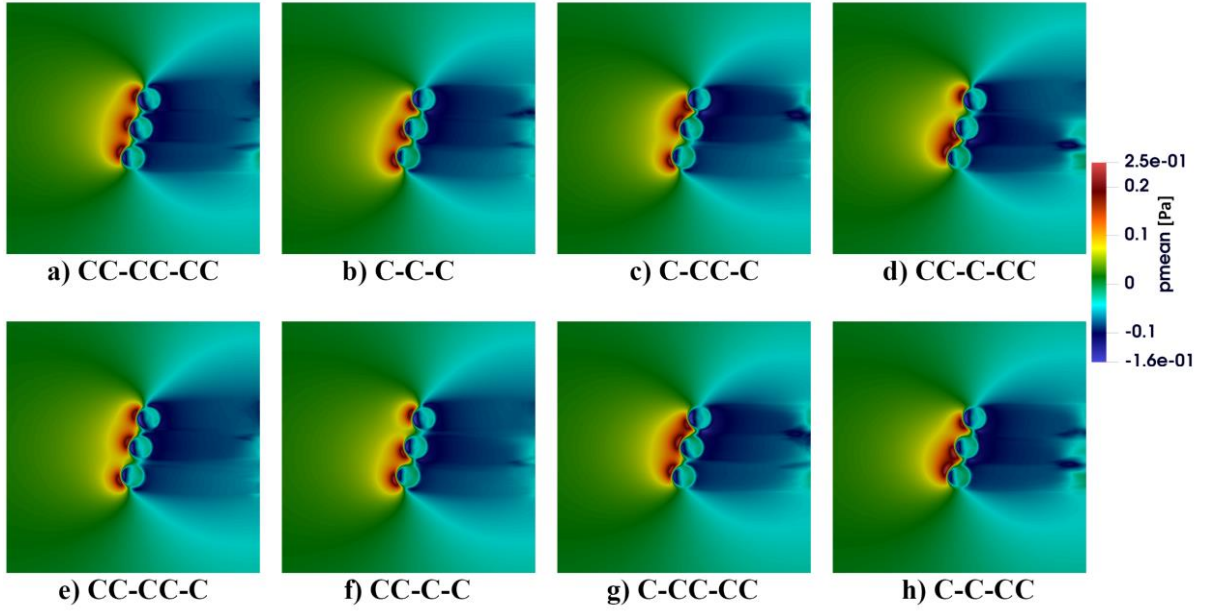

**FIG. 14.** Mean pressure contours for different staggered cases [The flow is from left to right]: a) CC-CC-CC, b)C-C-C, c) C-CC-C, d) CC-C-CC, e) CC-CC-C, f) CC-C-C, g) C-CC-CC, h) C-C-CC

This disparity in proximity affects the flow acceleration, which is a primary factor contributing to the performance enhancement in wind turbine clusters. The velocity upstream of the turbines undergoes a reduction due to the back effect created due to the rotation of the downwind turbines T2 and T1. Fig. 15 (a) compares the upstream velocity in front of turbine T3 for two different configurations: the staggered CC-CC-CC and C-C-C layouts. In the clockwise rotation of the intermediate turbine T2 case, the available kinetic energy for turbine T3 is reduced compared to the counter-clockwise rotation of the turbine T2 case. This decrease can be attributed to the rotation of the intermediate turbine T2, which creates a high-pressure region that affects the flow between T3 and T2. Since most power output originates from the turbine's upwind path, the intermediate turbine's clockwise rotation decreases the performance of the upwind turbine.

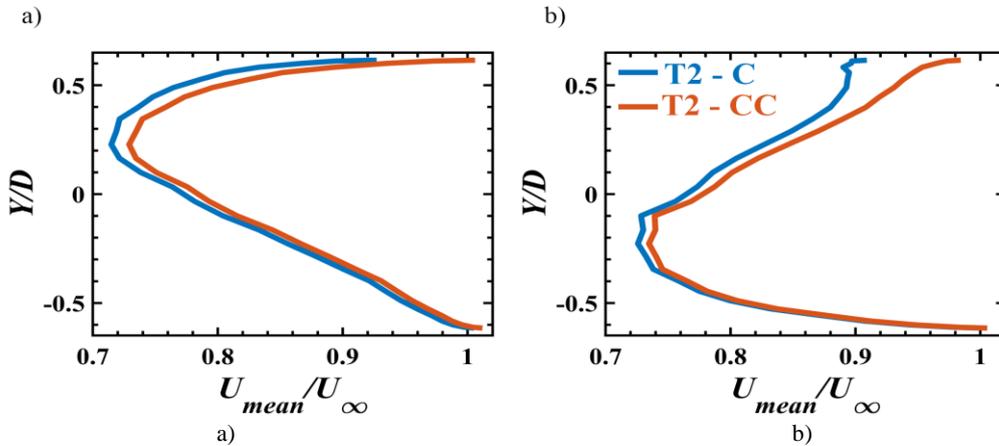

**FIG. 15.** Distribution of mean velocity on the upwind side of turbine T3 (located at 14.52D downstream from the inlet): a) Counter Clockwise b) Clockwise for two different rotations of the downstream turbine T2.

The contour plots (in Fig. 14) reveal that, due to the clockwise rotation of T2, high-pressure regions are formed near the right half of the upwind path of turbine T3. The same trend is observed for turbine T2, which results in decreased performance for both turbines, leading to an overall decrement in the performance of the turbine cluster. In the clockwise co-rotating inline configuration (C-C-C) scenario, the influence of symmetry effects becomes evident (see Fig. 16), eliminating the back effect caused by rotation. This is due to all turbines being aligned in a straight line. Consequently, the performance of the C-C-C case is equivalent to that of the CC-CC-CC configuration in the inline arrangement. Symmetry allows T3 in C-C-C to function as T1 in CC-CC-CC; similarly, T1 in C-C-C acts as T3 in CC-CC-CC. This symmetry exchange ensures that the behavior and performance of the cluster remain consistent regardless of their arrangement within the inline configuration.

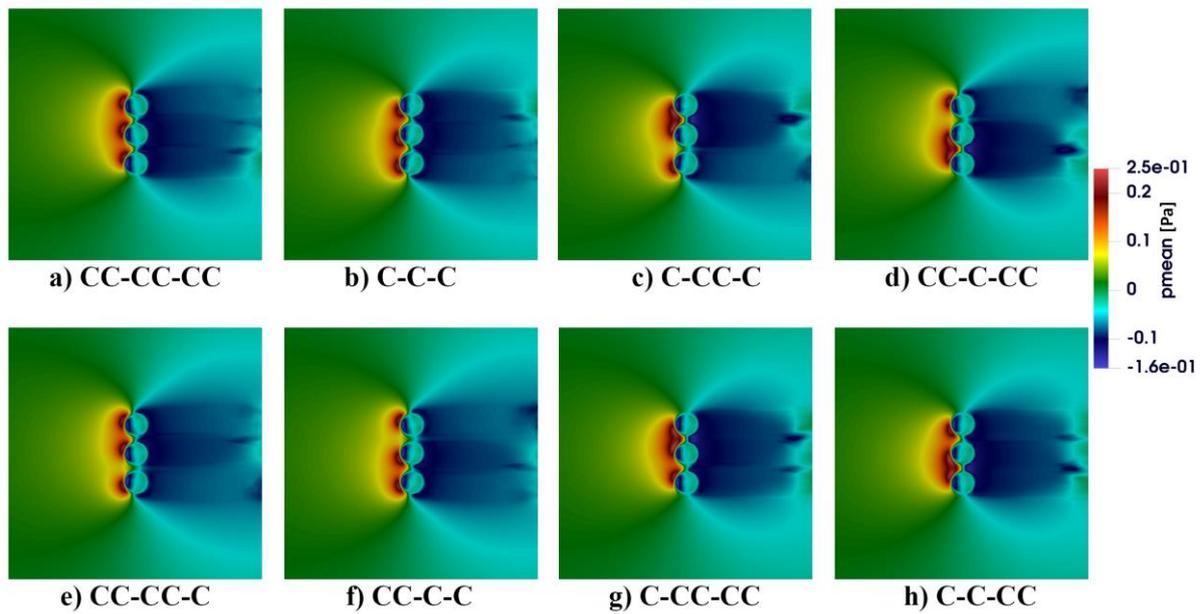

**FIG. 16.** Mean pressure contours for different inline cases: a) CC-CC-CC, b)C-C-C, c) C-CC-C, d)CC-C-CC, e) CC-CC-C, f) CC-C-C, g) C-CC-CC, h) C-C-CC

**Table IX.** Symmetric cases in the inline configuration

| Case | $C_p / C_{p_{iso}}$ | Symmetry case |
|---|---|---|
| CC-CC-CC | 1.093 | C-C-C |
| C-CC-C | 1.102 | CC-C-CC |
| CC-CC-C | 1.095 | CC-C-C |
| C-CC-CC | 1.10 | C-C-CC |

By comparing the performance of staggered and inline cases, a notable observation emerges: except for the anti-clockwise co-rotating staggered case (CC-CC-CC), there is a consistent decline in performance across all other cases, as shown in Table VII. In certain inline cases, the performances are identical, demonstrating symmetry attributable to the considered configuration and boundary conditions. Table IX provides an overview of these specific cases and their corresponding performances.

Upon careful examination of the staggered and inline configurations, it becomes evident that there is a decrease in performance from the CC-CC-CC case to the C-CC-C case in the staggered configuration. Conversely, in the inline configuration, the performance shows an opposite trend, increasing from the CC-CC-CC case to the C-CC-C case. To further illustrate this, Fig. 16 displays the mean pressure contour plots for different rotation combinations in the inline configuration. Interestingly, in the inline configuration, there is no noticeable back effect that could potentially hinder the performance of the upwind turbine. This absence of a detrimental back effect contributes to the sustained performance and efficiency of the inline configuration.

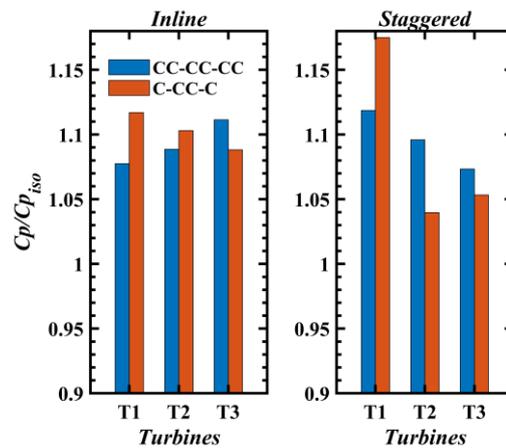

**FIG. 17.** Performance comparison for the selected four cases

Fig. 17 compares the individual turbine performances in the staggered and inline configurations, considering both co-rotating and counter-rotating cases. Notably, turbine T1 improves performance from the CC-CC-CC case to the C-CC-C case, irrespective of the configuration. The clockwise rotation increases energy extraction from the flow field, as depicted in Fig. 18 (a).

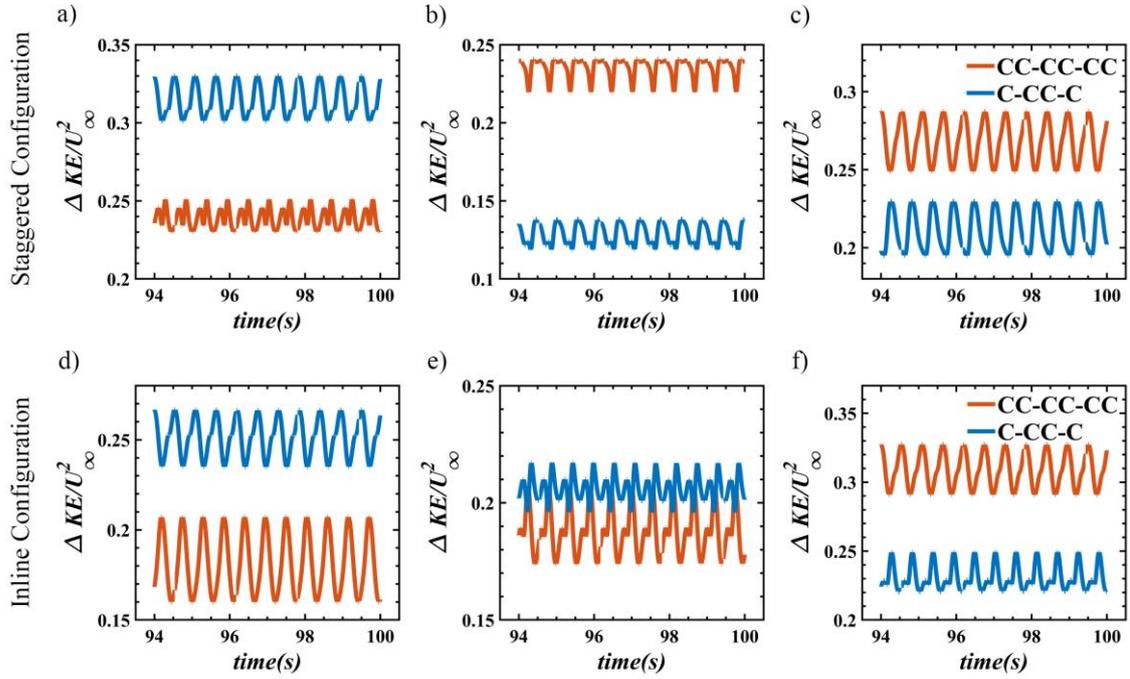

**FIG. 18.** Kinetic energy change in a staggered configuration (first row): a) turbine T1 b) turbine T2 c) turbine T3, and in inline configuration (second row): d) turbine T1 e) turbine T2 f) turbine T3

Several observations can be made by analyzing the contour plots of the mean pressure distribution in Fig.14 (a) and (c). For the clockwise rotating turbine T1, the high-pressure region is predominantly confined to the lower half of the upwind path, while the remaining half exhibits a relatively uniform pressure distribution. In contrast, the counterclockwise rotating turbine T1 displays a more evenly distributed high-pressure region along the upwind path. This disparity results in a reduction in the available energy content.

Fig. 19 (a) illustrates that power generation from a single blade is highest in the upwind path compared to the downwind path. Consequently, the performance improvement is more pronounced in the second half of the upwind path for the clockwise rotating T1, as opposed to the fully blocked counterclockwise turbine T1. Due to the uneven pressure distribution over the upwind path for turbine T1 in both clockwise and counterclockwise rotations, a noticeable phase shift in the power coefficient can be observed in Fig. 19 (a), with the isolated turbine's power coefficient lying between the two cases. The phase shift towards the left in the clockwise rotating turbine T1 enhances its performance.

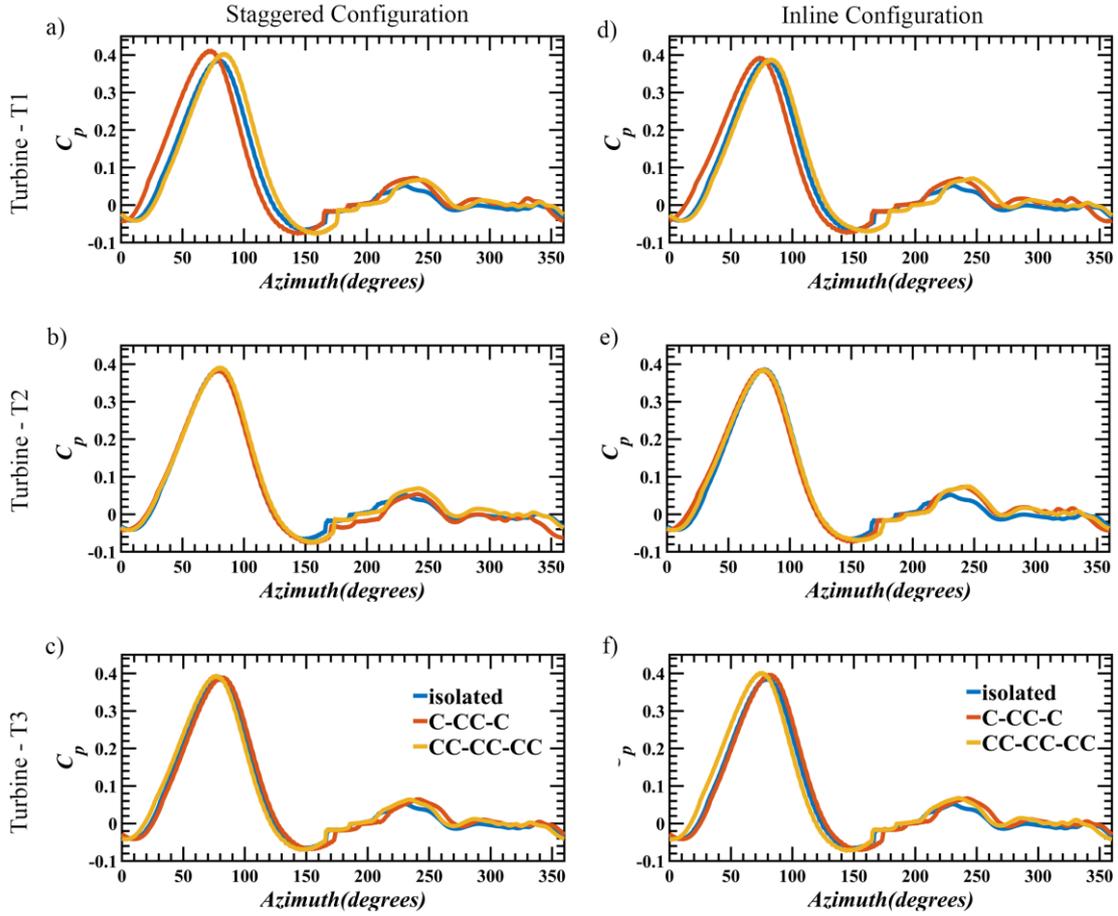

**FIG. 19.** Cp variation over Azimuth of the turbine for a single blade in a staggered configuration (left column): a) turbine T1, b) turbine T2, c) turbine T3, and in inline configuration (right column): d) turbine T1, e) turbine T2, f) turbine T3.

The same principle applies to the inline case, although the performance increment is more significant in the staggered configuration compared to the inline configuration. This is because the downwind position of turbine T1 in the staggered configuration provides it with greater energy availability than its position in the inline configuration, as shown in Fig. 10 (a). Turbine T3 decreases performance when transitioning from counterclockwise to clockwise rotation, as observed in the staggered and inline configurations. This phenomenon can be explained similarly to the analysis conducted for turbine T1, based on the mean pressure contour plots depicted in Figs. 14 (a) and (c). In the case of counterclockwise rotation, the upper half of the turbine experiences high-pressure regions, while the lower half exhibits nominal pressure. Conversely, for the clockwise rotation of turbine T3, the entire upwind path is enveloped by high-pressure regions. As mentioned earlier, this difference in pressure distribution results in improved performance for the counterclockwise rotating turbine T3, specifically in its lower half. These conclusions also hold for the inline configuration, and the enhanced performance can be attributed to the absence of a downstream turbine, which distinguishes it from the staggered configuration. The comparison of kinetic energy changes, as depicted in Figs. 18 (c) and (f) illustrate that the counterclockwise rotating turbine extracts more energy from the flow field than the clockwise rotating turbine.

An intriguing observation emerges when comparing the performance of turbine T2 in the C-CC-C case to that in the CC-CC-CC case within the staggered configuration, despite both cases involving counter-clockwise rotation of T2. Surprisingly, there is a decrement in performance for T2 in the C-CC-C case. However, in the inline configuration, the opposite trend is observed. This discrepancy explains the overall performance enhancement of the entire cluster in the counter-rotating inline configuration compared to the staggered configuration. As previously discussed, when a downstream turbine rotates clockwise in the staggered configuration, it adversely affects the upstream turbine, regardless of its rotation direction. In the C-CC-C staggered configuration, turbine T1 rotates clockwise, leading to a detrimental impact on the available kinetic energy in front of turbine T2. Consequently, T2's performance is reduced. In contrast, in the CC-CC-CC staggered configuration, the downstream turbine rotates counter-clockwise, positively influencing turbine T2, thus enhancing its performance.

This phenomenon is depicted in the kinetic energy change plot illustrated in Fig. 18 (b). The contrasting effects of the downstream turbines' rotation directions on the performance of T2 further emphasize the significance of configuration and rotation interactions in determining the overall efficiency of the wind turbine cluster.

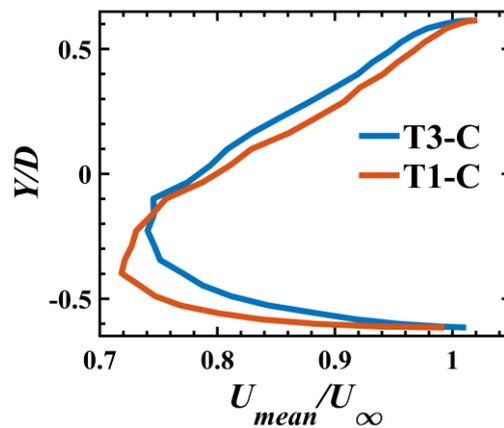

**FIG. 20.** Mean Velocity in the upwind path for turbines T1 and T3 in C-CC-C case in an inline configuration.

In the case of the inline configuration, an intriguing phenomenon occurs whereby counter-forward rotating turbine pairs exhibit enhanced performance, as also documented by Su et al.[24]. This phenomenon can be attributed to the formation of high-pressure regions in only one half of the upwind path of the turbine as shown in Fig. 16 (c). Consequently, the turbine can extract more energy from the flow in the other half, resulting in an overall improvement in the performance of the turbine cluster, as shown in Fig. 20, where there is an increment in upwind velocity observed in the clockwise rotating turbine T1. So, the performance of turbine pairs T1 and T2 (C-CC) is greater compared to the pair T2 and T3 (CC-C). The counter-rotating pair is more advantageous in extracting energy from the flow than the co-rotating pair. Both turbines extract more energy from the accelerated flow in the counter-forward rotating pair. In contrast, in the co-rotating pair, only one of the turbines can harness energy from the flow. By considering the analogies mentioned above, we can conclude the variations in performance across different configurations.

Now, let's examine the reasons for the decrease in performance in the other cases compared to the CC-CC-CC case in the staggered configuration. In the CC-C-CC and CC-C-C cases, the rotation of the intermediate turbine (T2) in a clockwise direction has a negative impact on the upwind turbine, as discussed earlier, leading to a reduction in the performance of turbine T3. Similarly, in the CC-CC-C case, due to the clockwise rotation of turbine T3, a high-pressure region accumulates along the upwind path, decreasing its performance. In the C-CC-CC and C-C-CC cases, the same effect described in the CC-C-CC case occurs, but the impact is on the intermediate turbine T2 and turbine T3, respectively.

Consequently, the inline configuration, where turbine T1 and T3 rotate clockwise, has been observed. In contrast, T2 rotates counter-clockwise (C-CC-C) and exhibits enhanced performance compared to the best-performing staggered counter-clockwise co-rotating case (CC-CC-CC). This configuration demonstrates improved performance and occupies a smaller area, aligning with the ultimate objective of this analysis. The findings suggest that the inline configuration balances performance enhancement and spatial efficiency, making it a good choice.

## IV. CONCLUSION

The present study adds several adjustments to incorporate specific characteristics and phenomena associated with VAWTs into the open-source solver. Notably, the unsteady effects, such as dynamic stall modeling and added mass effects (tailored to the unique features of VAWTs), the flow curvature modeling, and the end effects modeling are not part of the original open-source code. Overall, these advancements in the code provide a robust platform for studying and analyzing VAWTs, allowing for a comprehensive understanding of their performance and characteristics. Extensive validation studies are performed to verify the correctness of the modeling, and the results exhibit a strong agreement with both experimental and numerical data available.

Further, the line configuration proposed by Zhang et al.[22] is initially considered, and the impact of different spacing values on the cluster's performance has been investigated. To gather data, probes are strategically placed around the turbines, with additional probes positioned 1D ahead and 1D behind the three turbines. The configuration featuring a streamwise separation ($X_{sep}$) of 0.34D and a transverse separation ($Y_{sep}$) of 2.5D demonstrates superior performance to all other streamwise and transverse spacings combinations. The performance of turbine T1 improves due to the increased energy availability for T1 as the streamwise separation increases. In contrast, the performance of turbine T3 decreases due to the influence of back effects. The intermediate turbine T2 performance behavior resembles the cluster and determines the optimal performance configuration for the given stream-wise and transverse spacing combinations.

Finally, the impact of individual turbine rotation on the performance of clusters has been investigated for two distinct configurations: staggered and inline. In the staggered configuration, the downstream turbine's rotation direction significantly impacts the upstream turbine's performance. Specifically, a clockwise rotating downstream turbine diminishes the performance of the upstream turbine, regardless of its rotation. Conversely, a counter-clockwise rotating downstream turbine enhances the performance of the upstream turbine. Counter-rotating configurations outperform the co-rotating setups in the inline configuration. Counter-rotation analysis reveals that reducing streamwise separation makes it possible to align the turbines inline without sacrificing performance. This inline alignment boosts the power density of the turbine cluster, which aligns with the main purpose of the study.

## ACKNOWLEDGMENTS

The authors would like to acknowledge the National Supercomputing Mission (NSM) for providing the computational resources of 'PARAM Sanganak' at IIT Kanpur, which is implemented by C-DAC and supported by the Ministry of Electronics and Information Technology (MEITy) and Department of Science and Technology (DST), Government of India. The authors would also like to acknowledge the IIT-K Computer Center (www.iitk.ac.in/cc) for providing the resources to perform the computation work.

## AUTHOR DECLARATIONS

**CONFLICT OF INTEREST**

The authors have no conflicts to disclose.

**DATA AVAILABILITY**

The data that support the findings of this study are available from the corresponding author upon reasonable request.

**APPENDIX: FLOW CHART OF THE ALM IMPLEMENTATION**

Figure 21 below explains the implementation of the ALM technique into the code for computations of the VAWTs.

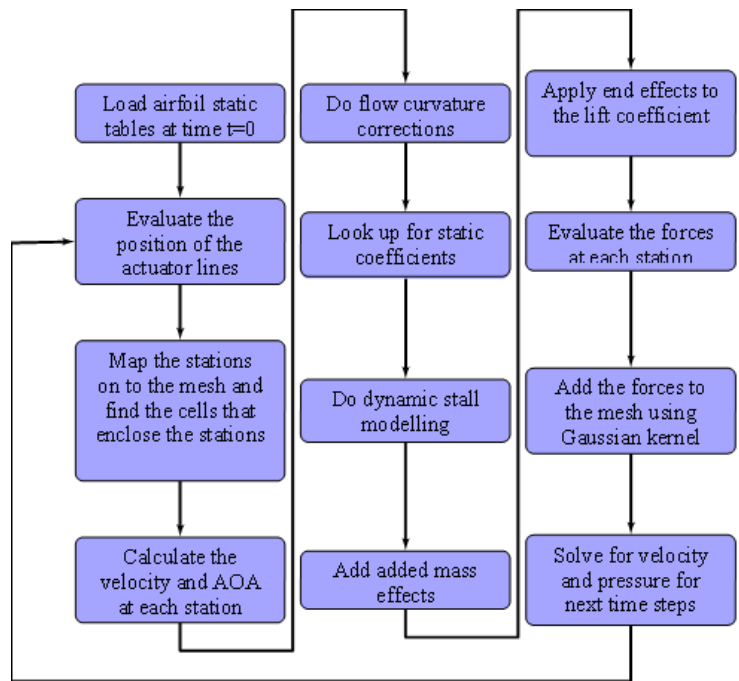

**FIG. 21.** Flow chart for the ALM implemented in the code